# On the relationship between optical and x-ray luminosity of quasars

Fabio La Franca[1,2]*, Alberto Franceschini[3], Stefano Cristiani[1], and Roberto Vio[1,4]

[1] Dipartimento di Astronomia dell'Università di Padova, Vicolo dell'Osservatorio 5, I-35122 Padova, Italy
[2] European Southern Observatory, Karl Schwarzschild Straße 2, D-85748, Garching bei München, Germany
[3] Osservatorio Astronomico di Padova, Vicolo dell'Osservatorio 5, I-35122 Padova, Italy
[4] ESA, IUE Observatory, Villafranca del Castillo, Apartado 50727, 28080 Madrid, Spain

**Abstract.** The issue of the X-ray to optical luminosity relationship ($L_o - L_x$) is addressed for both optically and X-ray selected quasar samples. We have applied a generalized regression algorithm for the case of samples involving censored data, with errors on both the dependent and the independent variable. Contrary to some previous results, we find that such relationship is consistent with being a linear one ($L_x \propto L_o$). We argue that previous reports of non-linear relationships (i.e. $L_x \propto L_o^e$ with $e < 1$) are due to the neglect of the influence of the photometric errors, the precise knowledge of which strongly influences the reliability of the results. Further progresses in the determination of the $L_o - L_x$ relationship can be achieved with ROSAT observations of the new generation of large bright quasar surveys.

**Key words:** quasars: general – X-rays: galaxies

## 1. Introduction

The study of the relationship between optical ($L_o$) and soft X-ray ($L_x$) luminosities of quasars is needed to relate the quasar statistics (luminosity functions and evolution) in the two wavebands, and to better define the quasar's broad-band emission. So far the bulk of the available X-ray flux data has come from the *Einstein* mission. All comparative analyses of the evolution in the two bands have shown that the evolution rate is slower in the X-ray than in the optical (e.g. Maccacaro, Gioia & Stocke 1984). Independently but consistently, a non-linear relationship between X-ray and optical luminosity of the type

$$L_x \propto L_o^e, \qquad (1)$$

*Send offprint requests to*: F. La Franca
\* *e-mail*: lafranca@astrpd.pd.astro.it

with $e \sim 0.8$, has been inferred by Kriss & Canizares 1985, Avni & Tananbaum 1986, Della Ceca et al. 1992, Boyle et al. 1993. Recently Wilkes et al. (1994) have studied the $L_o - L_x$ relationship in a sample of optically selected quasars observed by the IPC on board of *Einstein*, and have confirmed the previous result by Avny & Tananbaum (1986) of a non linear relationship between optical and X-ray luminosities.

These questions have been reconsidered by Franceschini et al. (1994), who showed that the evolution rate of X-ray selected AGNs could have been underestimated and actually be comparable to that of the quasars in the optical, with a linear scaling between the emission in the two bands.

In this paper we explore in more detail the origin of the discrepancies on the $L_o - L_x$ relationship, and discuss in particular the statistical uncertainties in the determination of the regression parameters for both optical and X-ray selected samples. We finally discuss the relevance of combining the ROSAT All Sky survey data with the new bright optical quasar samples, having a better photometric quality than previous samples. We adopt throughout the values $H_o = 50\ Km\ s^{-1}Mpc^{-1}$, $q_o = 0$ for the Hubble and deceleration parameters.

## 2. Generalized regression analyses

Studying the relationship between the X-ray and optical luminosity of quasars, with the data presently available in the literature, is mainly a problem of finding the correct regression algorithm and carefully determining the photometric errors.

Different results on the relationship between two quantities are usually obtained from regression analyses as a consequence of which one is chosen as the dependent and which one as the independent variable. The simplest least-squares regression assumes that the independent variable is something given *a priori* and not subject to any obser-realistically take into account the fact that usually (as in our case) data points are affected by errors in both $x$ and $y$ coordinates. The slope of the best-fit straight line obtained in this way is systematically lower than the correct one.

Two significantly different situations have to be distinguished: X-ray selected samples or optically selected samples.

In the first case, thanks to the potentially unlimited amount of observation time with ground-based telescopes, the optical follow-up is in practice complete, i.e. the optical data for the X-ray sources are not affected by *censoring*. Various statistical methods have been suggested to solve the problem in such a case (see e.g. Isobe et al., 1990). In the following we will adopt a procedure of generalized orthogonal regression, taking into account both measurements errors and intrinsic variances (i.e. scatter around the regression line unaccounted for by the observational uncertainties) in the data (see also Fasano & Vio 1988). The problem is to estimate the parameters of the linear regression $Y = a + bX$, when also the dependent variable is subject to measurements errors.

In terms of the first two empirical moments of the joint distribution $F(x, y)$ of the data $\{x\}$ and $\{y\}$ we have:

$$b = \frac{cov(x,y)}{\sigma_x^2}, \qquad (2)$$

$$a = \bar{y} - b\bar{x}. \qquad (3)$$

Here $\bar{x}$, $\bar{y}$ are the sample means, $cov(x, y)$ is the covariance and $\sigma_x^2$ is the variance of the distribution of the $x$ variable:

$$cov(x,y) = <(x - \bar{x})(y - \bar{y})>,$$

$$\sigma_x^2 = <(x - \bar{x})^2>.$$

If also the dependent variable is subject to experimental errors, the estimator of the slope in eq. (2) is biased toward a lower value. Let's define $e_x$ and $e_y$ the errors in the $x$ and $y$ axes, and $\sigma_{e_x}$ and $\sigma_{e_y}$ their standard deviations, with $\sigma_{e_x} \neq \sigma_{e_y}$ but constant for all the data. In this case $\sigma_x^2 = \hat{\sigma}_x^2 + \sigma_{e_x}^2$ where $\hat{\sigma}_x^2$ is the "true" variance of the distribution of $x$ and $\sigma_{e_x}^2$ is the contribution due to the measurement errors. Since the expected value of $cov(x,y)$ is not affected by the measurement errors of $x$, the larger is $\sigma_{e_x}^2$ the lower will be the expected value of $b$. If $b$ is biased, according to eq. (3) also the estimator $a$ will be biased.

Assuming gaussian errors, we can associate to each observed point $(x_i, y_i)$ a *bivariate normal* probability density function $p_i(x, y)$, whose characteristic variances along the axes are $\sigma_{e_x}^2$ and $\sigma_{e_y}^2$. In this case, the regions around each observed point, where the "true" point can be found with some given probabilities (confidence regions), are bounded by the families of ellipses

$$\frac{(x-x_i)^2}{\sigma_{e_x}^2} + \frac{(y-y_i)^2}{\sigma_{e_y}^2} = k_i^2 \qquad (4)$$

at varying $k_i$. The point of the true straight line that has most likely generated a given observed point is that tangential to the family (4). The tangency condition is found to be

$$k_i^2 = (y_i - a - bx_i)^2 / (\sigma_{e_y}^2 + b^2 \sigma_{e_x}^2).$$

According to the Maximum Likelihood principle, the most likely straight line is obtained by maximizing the product of the probabilities $p_i(x, y)$ computed at the tangential points, that is by minimizing the sum

$$L^2 = \sum_i k_i^2. \qquad (5)$$

The minimization of quantity (5) with respect to $a$ and $b$ provides the estimator in eq. (3) and the equation

$$b^2 + b\frac{(\sigma_{e_y}/\sigma_{e_x})^2 \sigma_x^2 - \sigma_y^2}{cov(x,y)} - (\frac{\sigma_{e_y}}{\sigma_{e_x}})^2 = 0. \qquad (6)$$

It is possible to show that the same results are obtainable by following a Least Squares approach (see Press & Teukolsky, 1992).

In the case of optically selected samples, because of the limits in the duration of X-ray missions, in addition to the uncertainties due to the errors on both axes, the presence of upper limits to the X-ray flux has to be taken into account. Avni et al. (1980) and Avni e Tananbaum (1986) have developed a regression method which deals with upper limits, that is based on the principles of the "Detections and Bounds" (DB) method, but does not consider the influence of errors on both axes. For the most generalized cases, in which both upper limits and errors in the $x$ and $y$ axes are present, we have used a simple extension of the Schmitt's technique (Schmitt, 1985). Note that other linear regression methods are available (for a review see Isobe, Feigelson & Nelson 1986), but their extension to the case of errors in both axes is not straightforward.

In its original formulation, the Schmitt's method provides an estimate of the parameters of the linear regression $Y = a + bX$, computing the estimators (2) and (3) by taking into account the presence of upper limits in the computation of $\bar{x}$, $\bar{y}$, $cov(x, y)$, $\sigma_x^2$. It is then straightforward to take into account the presence of errors on both axes by substituting the estimator (2) with estimator (6), and by computing the mean values $\bar{x}$, $\bar{y}$, the covariance and the variance of distribution of the $x$ variable, exactly as explained by Schmitt (1985).

# AGNs

The $L_o-L_x$ relationship for X-ray selected AGNs has been extensively investigated by Franceschini et al. (1994). We outline here the main results. The sample used is the combination of the EMSS (Stocke et al. 1991) with the deep ROSAT observations from Boyle at al. 1993, resulting in the largest sample available at present.

A generalized regression analysis allowing for errors and intrinsic variances in both $M_B$ and $L_x$ has been performed as explained in Section 2. We have assumed that the *Einstein* X-ray fluxes have uncertainties of 30%, a value somewhat larger than reported by Gioia et al. (1990), but realistically closer to account for various additional uncertainties, in particular those due to the effect of a variable galactic low energy absorption. An error of 0.5 magnitudes on $M_B$ has also been assumed. As explained below, this takes into account: 1) that the EMSS optical photometry is a collection of CCD data and Schmidt plate POSS measurements (with an average error of 0.4 magnitudes), 2) that the quasar variability adds an uncertainty of $\sim 0.2$ magnitudes, 3) and that an error of 0.15 magnitudes due to intrinsic spread on the optical spectral index should be taken into account.

The effect of the variability is difficult to address, as it significantly depends on the quasar absolute optical luminosity, and increases with decreasing redshift. From the analysis of the structure function (Hook et al. 1994) the typical amplitude of the quasar variability increases with the time-base in the quasar rest frame according to the formula:

$$<|\Delta m|> = [0.196 + 0.033(M_B + 25.7)](1 - e^{-\tau/0.38})^{1/2}, \quad (7)$$

where $\tau$ is expressed in years. Even if we take into account the redshift correction, it results that on average the optical photometry of the *Einstein* AGNs has been carried out some $\tau > 0.38(1+z)$ years later or before the X-ray measurements (at a typical redshift $z \sim 0.5$ for the EMSS), and we should expect an average error due to variability of 0.2 magnitudes for quasar having a typical absolute magnitude $M_B \sim -26$ (0.1 for $M_B \sim -30$ and 0.3 for $M_B \sim -23$).

The dispersion of the optical spectral index $\alpha$ ($F_\nu \propto \nu^{-\alpha}$) is typically 0.25-0.30 (e.g. Sargent, Steidel & Boksenberg, 1989), this figure originating an error of $\simeq 0.15$ magnitudes at an average redshift $\sim 1$, in agreement with what observed by La Franca, Cristiani & Barbieri (1992).

We have excluded from our analysis objects having $L_{x(0.3-3.5\ keV)} < 10^{44}\ erg\ s^{-1}$ and $z < 0.2$ in order to avoid the presence of spurious trends introduced by the contribution of the low-redshift, low-luminosity objects in which the emission from the host galaxy becomes non-negligible. It results that the functional dependence between the X-ray and optical luminosity for AGN at $z > 0.2$ and $L_x > 10^{44}$ is basically a linear one, with no indication tical to X-ray luminosity ratio. The best-fit regression for 268 AGNs is

$$M_B = -2.5(\pm 0.1)\ log L_{(0.3-3.5\ keV)} + 87.9(\pm 4.9), \quad (8)$$

or

$$log L_{(2\ keV)} = log L_{(2500\ \text{Å})} - 3.5, \quad (9)$$

(quoted uncertainties are bootstrap errors at 95% confidence). This corresponds to:

$$\alpha_{ox} = \frac{log(L_o/L_x)}{log(\nu_x/\nu_o)} = \frac{log(L_x/L_o)}{2.605} = 1.32, \quad (10)$$

$L_{(2\ keV)}$ and $log L_{(2500\ \text{Å})}$ being expressed in units of $erg/s/Hz$. An essentially linear relationship for medium-to-high luminosity objects is confirmed by a detailed analysis of the residual $M_B$ distribution. In particular, by adopting our best-regression solution, a Gaussian model with $\sigma(\Delta M_B) = 0.85$ fitted to the binned distribution of the residuals in eq. (8) provides a good fit (with a total $\chi^2 = 21$ for 22 independent bins).

## 4. The $L_o - L_x$ relationship for optically selected AGNs

The relationship between optical and X-ray emission for optically selected AGNs has been studied by various groups using data from the *Einstein* mission. Avni and Tananbaum (1986) using a sample with 94 X-ray detections and 60 upper limits found that the relationship between X-ray and optical luminosity was non-linear, with a best-fit value for the exponent in eq. 1 of $e \sim 0.8$. More recently and for a richer sample (179 detections and 164 upper limits), Wilkes *et al.* (1994) found a best-fit value for the exponent as small as $e \sim 0.7$.

We have verified that these dependences cannot apply to X-ray selected samples. Regressions lines with $e < 1$ entail residual distributions both markedly correlating with the independent variable $L_x$ and characterized by a significant skewness. If we adopt in particular $e = 0.7$ in eq. (1), we would obtain a distribution of the residuals which is clearly a function of the X-ray luminosity and to which a Gaussian model (even with a standard deviation increased to $\sigma(M) = 0.95$) would provide a very poor fit.

In order to better understand the inconsistencies between results of regression analyses of X-ray selected samples and those obtained for optically selected samples, we have analyzed the data of Wilkes et al. (1994) applying the method for DB data with errors on both axes as explained in Section 2. As discussed there, the results depend on the assumed errors of the measurements on the two axes. Unfortunately, the uncertainties on $L_x$ and $M_B$ are not known, but it seems reasonable to consider them comparable in the $log L_o - log L_x$ plane: an error of 30% in the fluxes as used for the *Einstein* data in the previous section corresponds to 0.75 magnitudes. Under this assumption we have obtained the relation

$$logL_{(2\ keV)} = 1.02(\pm 0.04)\ logL_{(2500\text{Å})} - 4.5(\pm 1.1), \quad (11)$$

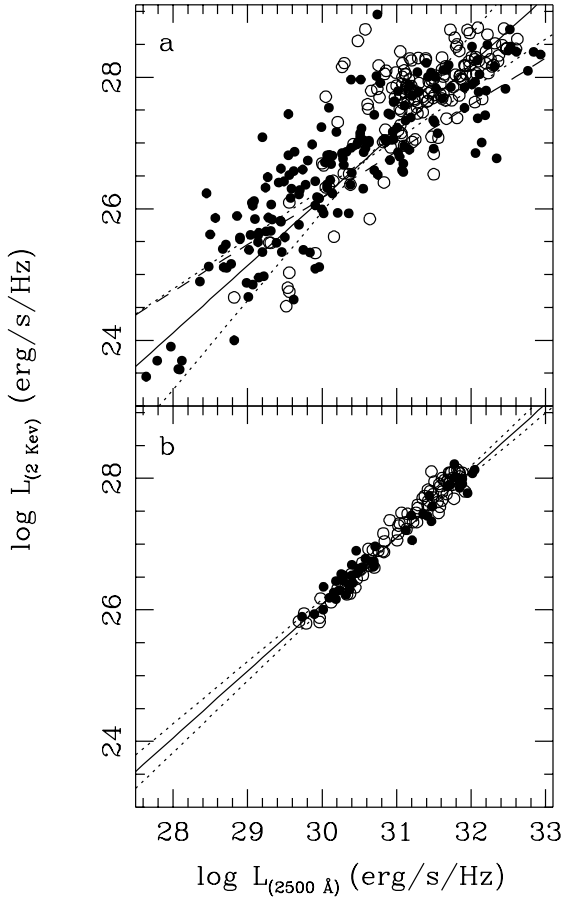

**Fig. 1.** Correlation between optical luminosity at 2500 Å and X-ray luminosity (2 keV) for optically selected samples. The filled circles indicate detections, the open circles indicate upper limits. (a) The sample of Wilkes et al. (1994): the continuous line is our best fit using the DB method, the dotted lines are the two extreme slopes in the case in which errors are assumed in turn only on the $x$ or on the $y$ axis; the dashed line is the best fit obtained by Wilkes et al. (1994). (b) Simulation of ROSAT observations (see text). 46 detections (filled circles) and 80 upper limits (open circles). The range in luminosities is smaller than in case a) because of the exclusion of the objects having $M_B > -23$.

(quoted uncertainties are bootstrap errors at 95% confidence) in agreement with the linear dependence between the two quantities found in Section 3 for the X-ray se- errors are assumed in the optical fluxes (this correspond to the result found by Wilkes et al. 1994) and 1.35 if no errors are assumed in the X-ray fluxes (see Figure 1a). In this case, following Wilkes et al. (1994), also the objects having $L_{x(0.3-3.5\ keV)} < 10^{44}\ erg\ s^{-1}$ have been included in the regression analysis.

As discussed by Cheng et al. (1984), X-ray flux limited observations will preferentially select objects with larger values of $L_x/L_o$, i.e. with smaller values of $\alpha_{ox}$. This explains the different constant terms in the $L_o - L_x$ relationships for X-ray selected (eq. [9]) and optically-selected quasars (eq. [11]). Let's indicate with $\alpha_{ox,X}$ and $\alpha_{ox,Optical}$ the spectral indices computed using X-ray and optically selected samples respectively, and with $\sigma_{\alpha,X}$ the standard deviation of the "observed" distribution of $\alpha_{ox,X}$. According to Cheng et al. (1984) and assuming a Gaussian distribution of the residuals ($\sigma_{M_B} = 0.85$, or $\sigma_{\alpha,X} = 0.153$), we obtain

$$\alpha_{ox,Optical} = \alpha_{ox,X} + 9\sigma^2_{\alpha,X} = 1.32 + 9(0.153)^2 = 1.53, \quad (12)$$

in good agreement with $\alpha_{ox,Optical} = 1.49$ obtained from eq. (11) at $logL_o = 31$.

## 5. Discussion and Conclusions

Our analysis shows that a linear relationship between optical and X-ray luminosity seems to be required by the available data from X-ray selected samples, and consistent with those on optical samples. Previous results favoring non-linear relationships may be due to the underestimation of the errors in the optical fluxes. Additional uncertainties are introduced by the large number of upper limits to the X-ray fluxes in the optical samples, by emission and/or extinction of the optical light from the host-galaxy in low-luminosity objects, and by the sparse nature of optical samples. From the above discussion it results clear that in order to unequivocally address the $L_o - L_x$ relationship it is necessary to analyze statistically-well-defined large data samples not only with an high level of photometric accuracy, but also with an accurate knowledge of the errors affecting the measurements on both the X-ray and optical fluxes. If the time distance between the measurements in the two bands is larger than few years, the variability adds errors of $\sim 0.1-0.3$ magnitudes. The achievement of accurate optical photometry provides little improvement if these effects are ignored.

Good opportunities in this sense are provided by recent advances: on the one hand the advent of the ROSAT satellite is providing new X-ray data, on the other hand three new bright optical quasar surveys are becoming available: the Edinburgh quasar survey (ROE, Goldschmidt et al. 1992); the Large Bright Quasar Survey ( LBQS, Hewett, Foltz & Chaffee 1993 and references therein); the Homoge-

1995).

We have analyzed the level of accuracy in the determination of the $L_o - L_x$ relationship that could be achieved from the ROSAT "All Sky Survey" data of quasars from two of these surveys: the HBQS and ROE samples, having almost the same color selection criterion, and the highest photometric quality. As for the analysis of the X-ray selected quasars in Section 3, we have excluded the fainter objects having $M_B > -23$, in order to avoid spurious trends introduced by the contribution of the light from the hosting galaxy, and the incompleteness which affects color selections for the intrinsically fainter quasars.

The HBQS and ROE surveys give a complete sample of about 120 quasars over an area of more than 800 deg$^2$ with $15 < B < 17.5$ (the inclusion of the LBQS would bring the sample size to roughly 200 quasars in this magnitude range). Particular care has been dedicated to the photometric accuracy (e.g. the correction of dis-uniformities in the response of the photographic material within a given Schmidt plate) reaching absolute errors less than 0.1 mag. To compute the total error on the optical fluxes which will be compared with the X-ray fluxes, we should also take into account the effect of variability (eq. 7), and of the dispersion of the optical spectral index as done in Section 3. For the HBQS there is a typical difference of about two years between the observations in the optical and ROSAT bands. The ROE survey has larger time differences. Assuming an average redshift of $z \sim 1$, according to eq. (7) we can estimate an error due to variability $\simeq 0.2$ magnitudes for both surveys. As discussed in Section 3, the dispersion of the optical spectral indices causes an error of $\simeq 0.15$ magnitudes at an average redshift $z \sim 1$. Then we expect a typical total uncertainty on the optical magnitudes (when reported at the time of the X-ray observations) of $\sigma_B \sim 0.25$ mag. The advantage in using samples such as the HBQS and ROE is to allow a well defined statistical definition and treatment of the photometric errors which represents a considerable improvement with respect to previous analyses which were based on incomplete samples from published inhomogeneous catalogs of quasars.

To estimate the errors on the X-ray measurements, a linear regression between the S/N ratio and the X-ray $F_{x(0.5-2\ keV)}$ ROSAT fluxes has been computed from a sparse sample of more than 200 quasars selected from the Véron & Véron (1993) catalogue and observed by the "All Sky Survey". It results:

$$log(\frac{S}{N}) = 0.53(\pm 0.02) log F_x + 7.09(\pm 0.08) \qquad (13)$$

(Molendi & Doublier, private communication). We have consequently used an "All Sky Survey" $3\sigma$ average detection limit of $F_{x(0.5-2.5\ keV)} = 3\ 10^{-13}\ erg\ cm^{-2} s^{-1}$, and have simulated the almost linear relationship between $L_x$ and $L_o$ as in eq. (11) for the quasars of the HBQS and ROE samples. Eq. (13) gives an average error HBQS and ROE surveys.

A plot of $L_x$ versus $L_o$ for simulated ROSAT observations of the HBQS and ROE quasars appears in Fig. 1b. It results that roughly one third of the quasars are expected to have a detection at a $3\sigma$ confidence level. Assuming $\sigma_B = 0.25$ and $\sigma_{log F_x} = 0.17$ constant for all the data, our method of DB allowing for errors on both axes reproduces the assumed value for the slope (eq. 11): $b = 1.02 \pm 0.03$. In the extreme cases in which the errors are assumed only either on the X-ray or on the optical fluxes, the slope value varies from 0.95 to 1.09. This is a quite narrower range than the interval $0.76 < e < 1.35$ previously obtained from the data collection by Wilkes et al. (1994), and shows the potential accuracy that could be achieved from this kind of observations.

Added to these uncertainties we should expect an intrinsic spread in the $L_o - L_x$ relationship, which at present is completely hidden by the the photometric errors in the data collected by Wilkes et al. (1994). Exploitation of the ROSAT "All Sky Survey" data for quasars in the HBQS and ROE samples will allow to test this idea.

The symmetric approach to the problem, based on optical identifications of ROSAT X-ray selected samples, is underway (Danziger et al., 1990, and Barcons et al., private communication). As demonstrated above, also in this case the reliability of the results will be strongly dependent on the precise knowledge of the photometric errors.

*Acknowledgements.* We tanks Silvano Molendi and Vanessa Doublier for providing us the relationship between the S/N ratio and X-ray fluxes from the ROSAT all sky survey. FLF acknowledges financial support from CNR. RV acknowledges receipt of an ESA fellowship. This work was partially supported by the ASI contract ASI-92-RS-102 and by the EC Human Capital and Mobility programme.